\documentclass[twocolumn,prl,amsmath,amssymb,showpacs,superscriptaddress]{revtex4-1}
\usepackage{epsf}      
\usepackage{graphicx}
\usepackage{color}
\usepackage{gensymb}
\usepackage{sidecap}

\begin{document}

\title {Tuning the metal-insulator transition in NdNiO$_3$ heterostructures via Fermi surface instability and spin-fluctuations}

\author{R. S. Dhaka}
\email{rsdhaka@physics.iitd.ac.in}
\affiliation{Swiss Light Source, Paul Scherrer Institute, CH-5232 Villigen PSI, Switzerland}
\affiliation{Institute of Condensed Matter Physics, Ecole Polytechnique F{\'e}d{\'e}rale de Lausanne (EPFL), CH-1015 Lausanne, Switzerland}
\affiliation{Department of Physics, Indian Institute of Technology Delhi, Hauz Khas, New Delhi-110016, India}
\author{Tanmoy Das}
\affiliation{Department of Physics, Indian Institute of Science, Bangalore-560012, India.}
\affiliation{Department of Physics, National University of Singapore, Singapore}
\author{N. C. Plumb}
\affiliation{Swiss Light Source, Paul Scherrer Institute, CH-5232 Villigen PSI, Switzerland}
\author{Z. Ristic}
\affiliation{Swiss Light Source, Paul Scherrer Institute, CH-5232 Villigen PSI, Switzerland}
\affiliation{Institute of Condensed Matter Physics, Ecole Polytechnique F{\'e}d{\'e}rale de Lausanne (EPFL), CH-1015 Lausanne, Switzerland}
\author{W. Kong}
\affiliation{Swiss Light Source, Paul Scherrer Institute, CH-5232 Villigen PSI, Switzerland}
\affiliation{Institute of Physics, Chinese Academy of Sciences, Beijing 100190, China}
\author{C. E. Matt}
\affiliation{Swiss Light Source, Paul Scherrer Institute, CH-5232 Villigen PSI, Switzerland}
\affiliation{Laboratory for Solid State Physics, ETH Z$\ddot{\rm u}$rich, CH-8093 Z$\ddot{\rm u}$rich, Switzerland}
\author{N. Xu}
\affiliation{Swiss Light Source, Paul Scherrer Institute, CH-5232 Villigen PSI, Switzerland}
\author{Kapildeb Dolui}
\affiliation{Department of Physics, National University of Singapore, Singapore}
\author{E. Razzoli}
\affiliation{Department of Physics, University of Fribourg, CH-1700 Fribourg, Switzerland}
\author{M. Medarde}
\affiliation{Laboratory for Developments and Methods, Paul Scherrer Institut, CH-5232 Villigen PSI, Switzerland}
\author{L. Patthey}
\affiliation{SwissFEL, Paul Scherrer Institute, CH-5232 Villigen PSI, Switzerland}
\author{M. Shi}
\affiliation{Swiss Light Source, Paul Scherrer Institute, CH-5232 Villigen PSI, Switzerland}
\author{M. Radovi{\'c}}
\email{milan.radovic@psi.ch}
\affiliation{Swiss Light Source, Paul Scherrer Institute, CH-5232 Villigen PSI, Switzerland}
\affiliation{SwissFEL, Paul Scherrer Institute, CH-5232 Villigen PSI, Switzerland}
\author{Jo$\ddot{\rm e}$l Mesot}
\email{joel.mesot@psi.ch}
\affiliation{Swiss Light Source, Paul Scherrer Institute, CH-5232 Villigen PSI, Switzerland}
\affiliation{Institute of Condensed Matter Physics, Ecole Polytechnique F{\'e}d{\'e}rale de Lausanne (EPFL), CH-1015 Lausanne, Switzerland}
\affiliation{Laboratory for Solid State Physics, ETH Z$\ddot{\rm u}$rich, CH-8093 Z$\ddot{\rm u}$rich, Switzerland}

\date{\today}                                         

\begin{abstract}
We employed {\it in-situ} pulsed laser deposition (PLD) and angle-resolved photoemission spectroscopy (ARPES) to investigate the mechanism of the metal-insulator transition (MIT) in NdNiO$_3$ (NNO) thin films, grown on NdGaO$_3$(110) and LaAlO$_3$(100) substrates. In the metallic phase, we observe three dimensional hole and electron Fermi surface (FS) pockets formed from strongly renormalized bands with well-defined quasiparticles. Upon cooling across the MIT in NNO/NGO sample, the quasiparticles lose coherence via a spectral weight transfer from near the Fermi level to localized states forming at higher binding energies. In the case of NNO/LAO, the bands are apparently shifted upward with an additional holelike pocket forming at the corner of the Brillouin zone. We find that the renormalization effects are strongly anisotropic and are stronger in NNO/NGO than NNO/LAO. Our study reveals that substrate-induced strain tunes the crystal field splitting, which changes the FS properties, nesting conditions, and spin-fluctuation strength, and thereby controls the MIT via the formation of an electronic order parameter with Q$_{AF}\sim$(1/4, 1/4, 1/4$\pm$$\delta$).

\end{abstract}

\maketitle

\section{\noindent ~Introduction}

The metal-insulator transitions (MITs) in complex oxides are often precursors to exotic ground states such as high temperature superconductivity, colossal magnetoresistance, and different types of charge-, spin- and orbital-ordered states \cite{Imada98,HwangNM12,Zubko}. In this context, rare earth nickelates (RNiO$_3$, where R = rare earth elements) can be viewed as model systems, which exhibit temperature-driven MITs \cite{Medarde97,Catalan}. Novel routes to engineering and controlling the MIT in thin films and heterostructures of RNiO$_3$ further offer the possibility to control the MIT with thickness via strain using different substrates \cite{Liu13,ScherwitzlPRL,Scherwitzl,CatalanPRB00,LiuPRL12,Stewart12,Benckiser11,Wu13}. Interestingly, the MIT in RNiO$_3$, which can be tuned by changing R \cite{Medarde97,Catalan}, by applying hydrostatic pressure \cite{Canfiled,Obradors}, or by epitaxial strain engineering in thin films \cite{Liu13,LiuPRL12,ScherwitzlPRL,Scherwitzl,CatalanPRB00,Kaiser11}, revealing that there are diverse possibilities for controlling correlation strength in these systems. 
 
The origin of the MIT in nickelate thin films is still unknown, and various explanations have been proposed, including magnetic order, charge order, orbital order, lattice distortions, or their combinations \cite{Munoz09,Scagnoli08,MedardePRB92,Kumar13,Scagnoli06,HanPRL11,ParkPRL12,LauPRL13}. Experimentally, it is reported that the MIT temperature coincides with the N\'eel temperature $T_N$ depending on the ionic radius of R \cite{Medarde97,Catalan}.  More recently, various groups have claimed that the MIT originates from the formation of charge-order due to NiO$_6$ octahedral distortions \cite{Johnston} or valence fluctuations between Ni$^{3+\delta}$ and Ni$^{3-\delta}$ ions \cite{StaubPRL02,Scagnoli05,MedardePRB09,Bilewska10}. An observed isotope effect supports this scenario \cite{MedardePRL98}. Furthermore, more recent transport measurements on NdNiO$_3$ thin films show non-Fermi-liquid behavior dominating in the samples which lie at the border of the onset of the magnetic ordering and MIT \cite{Liu13}. The result is suggestive of the presence of a `hidden' quantum critical point as a function of strain \cite{Liu13}. Interestingly, optical spectroscopies have identified that the MIT is associated with a spectral weight transfer from the mid-infrared region to a higher energy scale \cite{Stewart12,StewartPRL11}.

Recent ARPES studies of LaNiO$_3$ thin films show that the FS topology changes from three- to quasi two-dimensional with decreasing film thickness \cite{Yoo14,King14}. In addition, an observed FS superstructure in LNO films indicates some instabilities in the metallic phase \cite{Yoo14}. Based on these results, it is proposed that tunable charge order, spin-density-wave fluctuations, and dimensionality might be responsible for thickness dependent MIT in LaNiO$_3$ and PrNiO$_3$ films \cite{Yoo14,Hepting,Boris}. However, the origin of the possible FS instability and the mass renormalizations responsible for the insulating state in RNiO$_3$ are still highly debated \cite{HanPRL11,ParkPRL12,LauPRL13,StewartPRL11,Okazaki03,Schwier12,Vobornik99,Barman,MedardeEPL97}. In order to identify the microscopic origin of the MITs and the role of different orders, it is crucial to understand how spectral functions and electronic interactions evolve across the transition and for different substrates. Therefore, in this paper, we use high-resolution ARPES to directly map the evolution of the electronic structure across the MIT in NdNiO$_3$ (NNO) thin films grown on NdGaO$_3$ (NGO) and LaAlO$_3$ (LAO) substrates.

\section{\noindent ~Sample preparation, Characterization, and Methods}

{\bf Sample preparation.} The 20 unit cells (uc) of NNO thin films were grown on NGO and LAO single-crystal substrates by pulsed laser deposition (PLD). A sintered stoichiometric NdNiO$_3$ pellet was used as an ablation target. A Nb-doped yttrium-aluminum-garnet (Nd:YAG) laser was used in its frequency-quadruple mode (266~nm) at a repetition rate of 2~Hz and an energy of about 90~mJ/pulse. The substrates were heated by passing current through silicon wafers mounted underneath. The temperature of each substrate (monitored by an external pyrometer) was set to about 730$^o$C and the pressure of the oxygen was about 0.1~mbar during the deposition. After deposition, each sample was cooled down to room temperature while maintaining the same oxygen pressure for about one hour. 

\begin{figure}
\includegraphics[width=3.5in]{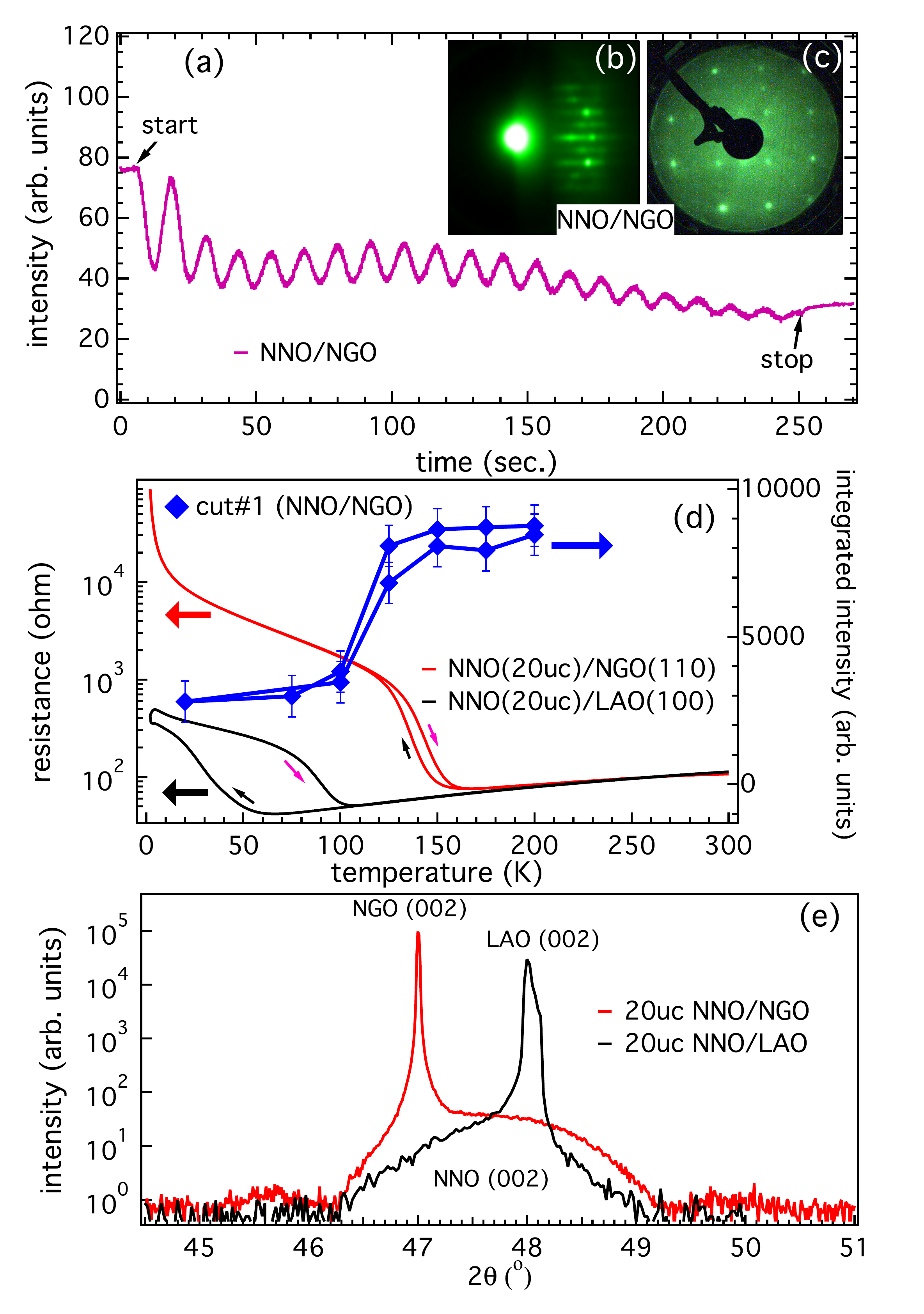}
\caption{Characterization of epitaxially grown NNO thin films on NGO(110) and LAO(100) substrates. (a) Typical RHEED intensity oscillations and (b) corresponding RHEED pattern of NNO/NGO. (c) LEED pattern of NNO/NGO taken at 211~eV electron energy. (d) Electrical resistance versus sample temperature during heating and cooling cycles, as well as the photoemission intensity of NNO/NGO (h$\nu=90$~eV) integrated near $E_{\rm F}$ (from -0.2 to 0.1~eV) plotted along cut$\#$1 (as the location is marked in the upper panel of Fig.~2a) on the right scale, (e) X-ray diffraction pattern ($\theta-2\theta$ scan) of NNO/NGO and NNO/LAO samples taken with Cu K$\alpha$ radiation showing the substrate peaks. The peaks from the NNO films are broad due to the very low thickness (only 20~uc).} 
\label{fig1}
\end{figure}

{\bf Characterization.} In Fig.~1 we summarize the structural and transport properties of the samples used in the ARPES experiments. Fig.~1(a) shows typical reflection high-energy electron diffraction (RHEED) oscillations during the growth of NNO thin films on NGO (for NNO/LAO, see Fig.~S4 of supplementary information), indicating epitaxial growth with the potential for unit-cell thickness control. The corresponding RHEED pattern is shown in Fig.~1(b). Based on the growth rate extracted from the RHEED oscillations, the films in our study have total thicknesses of about 20~uc (total deposition time of 240~s). In Fig.~1(c), a sharp diffraction pattern and some surface reconstruction derived spots are observed in a low-energy electron diffraction (LEED) image taken at 211~eV electron energy, confirming the well-ordered, high-quality single crystalline surfaces of the deposited thin films. The electrical resistance of the films was measured by the four-point van der Pauw method using a physical property measurement system. The resistance of 20~uc NNO films on NGO(110) and LAO(100) substrates (the actual samples on which ARPES measurements were performed) is shown in Fig.~1(d) as a function of temperature. The MIT of nominally $\sim140$~K is about 60~K lower than in the bulk NNO samples, consistent with previous reports \cite{Liu13,Scherwitzl}. As can be seen in Fig.~1(d), despite the large variation in the strain, a similar metallic behavior from 300~K down to 170~K is observed in samples grown on both the substrates. However, for NNO/NGO the resistance begins a step-like increase below about 155~K that extends until about 120~K. The resistance continues to increase gradually even at a lower temperature until a sharp increase in resistance is observed at around 20~K. No measurable hysteresis is observed down to 2~K, indicating a very sharp transition to a completely insulating state in this very low temperature range. On the other hand, resistance measurements of NNO/LAO films show a transition at around 60~K with broad hysteresis, which does not appear to saturate at the lowest temperature of 2~K. The temperature dependence of the resistance across the MIT, including its hysteretic behavior, correlates with the changes in quasiparticle spectral weight near the Fermi level seen in the ARPES data [Fig.~1(d)], which will be explained in more detail in the Results section.

 We further characterized the films using x-ray diffraction at room temperature with Cu K$\alpha$ radiation. Representative $\theta-2\theta$ scans of both the samples are shown in Fig.~1(e). No impurity phases have been detected and the position of the Bragg peaks are consistent with previous reports \cite{Liu13,Scherwitzl}. The diffraction peaks observed at about 47$^o$ and 48.2$^o$ are from (002) planes of the NGO and LAO substrates, respectively.  A broad peak centered at about 47.5$^o$ originates from the NNO thin films [as marked in Fig.~1(e)]. This broadening of the film peaks is expected because our films are only 20~uc thick. The full width at half maximum of the film is found to be about 0.15$^{\circ}$ in the rocking curve measurements (not shown), confirming the good quality of the NNO films (slight broadening may be due to the presence of dislocations \cite{Scherwitzl}). By comparing the in-plane pseudocubic lattice constants of the NGO ($a=3.858$~\AA) and LAO ($a=3.794$~\AA) substrates with NNO ($a=3.803$~\AA), the NNO films are nominally under tensile strain of +1.4\% and compressive strain of -0.3\% when deposited on NGO and LAO substrates, respectively.

{\bf ARPES measurements.} The freshly grown NdNiO$_3$ thin film samples were studied \emph{in-situ} using ARPES. The ARPES measurements were performed at the Surface/Interface Spectroscopy (SIS) X09LA beamline of the Swiss Light Source located at the Paul Scherrer Institute in Villigen, Switzerland \cite{Sassa}. The end station is equipped with a VG-Scienta R4000 electron energy analyzer and a six axis liquid He-cooled manipulator (CARVING). We used photon energies from 60~eV to 115~eV with both circular and linear photon polarizations. Spectra were acquired over a temperature range from 20 to 200~K. The energy and momentum resolutions were set to about 30~meV and $\approx0.009/0.019$~\AA$^{-1}$ (parallel/perpendicular to the analyzer slit), respectively. The binding energy scale was calibrated with a polycrystalline copper reference sample in direct electrical and thermal contact with the film. The base pressure of the UHV chamber during the measurements was below 5$\times$10$^{-11}$~mbar. Our results were reproduced on several thin film samples. 

{\bf Theoretical calculations.} The theoretical calculation of the complex self-energy effects is performed within the momentum dependent density fluctuations (MRDF) model \cite{AIPDas,PuDas}. We first derive a three-orbital tight-binding model to capture the essential non-interacting band structure. We include nickel $d_{x^2-y^2},~d_{z^2}$ orbitals whose related bands are observed to cross $E_F$ in all our samples, and also the $d_{xy}$ orbital, which crosses $E_F$ for the NNO/LAO case. The substrate dependent crystal field effect is parametrized by tunning the onsite potential difference of the $d_{xy}$ orbital with respect to its $e_g$ counterparts, as well as by changing the inter-orbital hopping between them. This causes not only that the $d_{xy}$ orbital shift toward to $E_F$, but also it helps to change the FS topology of the other two $e_g$ orbitals, in accordance with experiment. We note that the Ni $d_{xy}$ orbital is strongly hybridized with O $p$-orbitals, and thus we fit to the overall bandwidth. The tight-binding fitting to the density-functional theory band structure is given in the Supplementary Information.   

The electronic interactions in transition metal oxides are dominated by density-density correlations which include spin, charge and orbital fluctuations. The density fluctuation susceptibilities in the particle-hole channel represent the joint density of states (JDOS), which is calculated by convoluting the corresponding Green's function over the entire Brillouin zone (BZ). The many-body corrections are captured within the random-phase approximation (RPA) by including the onsite multiband components such as intra-, inter-orbital Hubbard interactions ($U$, $V$), Hund's coupling $J_H$, and the pair-change term $J^{\prime}$. Their values are found to be same for both samples with $U$=2~eV for $d_{x^2-y^2}$,1.5~eV for the $d_{z^2}$ orbital and 0.8~eV for the $d_{xy}$ orbital. The inter-orbital interaction is $V=$1.5~eV, Hund's coupling $J_H$=0.8~eV, and a pair-hoping term $J^{\prime}=$0.33~eV for all orbitals. 

We calculate the electronic self-energy due to the coupling of the density-fluctuations to the electronic states within the MRDF method \cite{AIPDas,PuDas,MoS2Das}.
\begin{eqnarray}
\Sigma_{mn,i}({\bf k},\omega) = \frac{1}{\Omega_{\rm BZ}}\sum_{{\bf q},st,\nu}\int_{-\infty}^{\infty} d\omega_p V_{mn,i}^{st}({\bf q},\omega_p)\Gamma_{mn,i}^{st,\nu}({\bf k},{\bf q})\nonumber\\
\times \left[\frac{1-f^{\nu}_{{\bf k}-{\bf q}} +n_p}{\omega+i\delta -\xi^{\nu}_{{\bf k}-{\bf q}} -\omega_p}
+\frac{f^{\nu}_{{\bf k}-{\bf q}} +n_p}{\omega+i\delta -\xi^{\nu}_{{\bf k}-{\bf q}} +\omega_p} \right].
\end{eqnarray} 
Here ${\bf k}$ and $\omega$ are the quasiparticle momentum and frequency, and ${\bf q}$ and $\omega_p$ are the bosonic excitation momentum and frequency, respectively. $m$, $n$, $s$, and $t$ are the orbital indices, and $\nu$ is the quasiparticle band index. The index $i$ stands for different types of fluctuatuion such as spin, charge, etc. $\Omega_{\rm BZ}$ is the electronic phase space volume. $f^{\nu}_{\bf k}$ and $n_p$ are the fermion and boson occupation numbers, and $\xi_{\bf k}^{\nu}$ is the corresponding tight-binding bands. The vertex correction $\Gamma_{mn,i}^{st,\nu}({\bf k},{\bf q})$ encodes both the angular and dynamical parts of the vertex, which are combined to obtain $\Gamma_{mn,\nu}^{st}({\bf k},{\bf q})=\phi^{\nu\dag}_{{\bf k}-{\bf q},s}\phi^{\nu}_{{\bf k}-{\bf q},t}(1-\partial \Sigma_{mn,i}^{\prime}({\bf k}-{\bf q},\omega)/\partial \omega)_{0}$, where $\phi$ is the eigenvector.  $V_{mn,i}^{st}({\bf q},\omega_q)$ is the fluctuation-exchange potential generated by the exchange of  spin, valence (charge), and orbital between different electrons: $V_{mn,i}^{st}({\bf q},\omega_p) = (\eta_{i}/2)\big[\tilde{U}_i\tilde{\chi}_{i}^{\prime\prime}({\bf q},\omega_p)\tilde{U}_i \big]_{mn}^{st}$, where $i$ stands for spin and charge (and orbital) components with $\eta_i$ = 3, 1, respectively. $\tilde{U}_i$ are the Coulomb interaction matrices, defined in the orbital basis, whose components are $U$, $V$, $J_H$, and $J^{\prime}$ as defined in the supplementary material. $\tilde{\chi}_i$ are the RPA spin and charge susceptibilities. Finally, the self-energy dressed  Green's function becomes $\tilde{G}^{-1}=\tilde{G}_0^{-1}-\tilde{\Sigma}$, where $\Sigma$ is the total self-energy matrix. We compute all self-energies self-consistently, until the total self-energy converges, as described in Refs.~\cite{AIPDas,MoS2Das,PuDas}.

\begin{figure}
\includegraphics[width=3.4in]{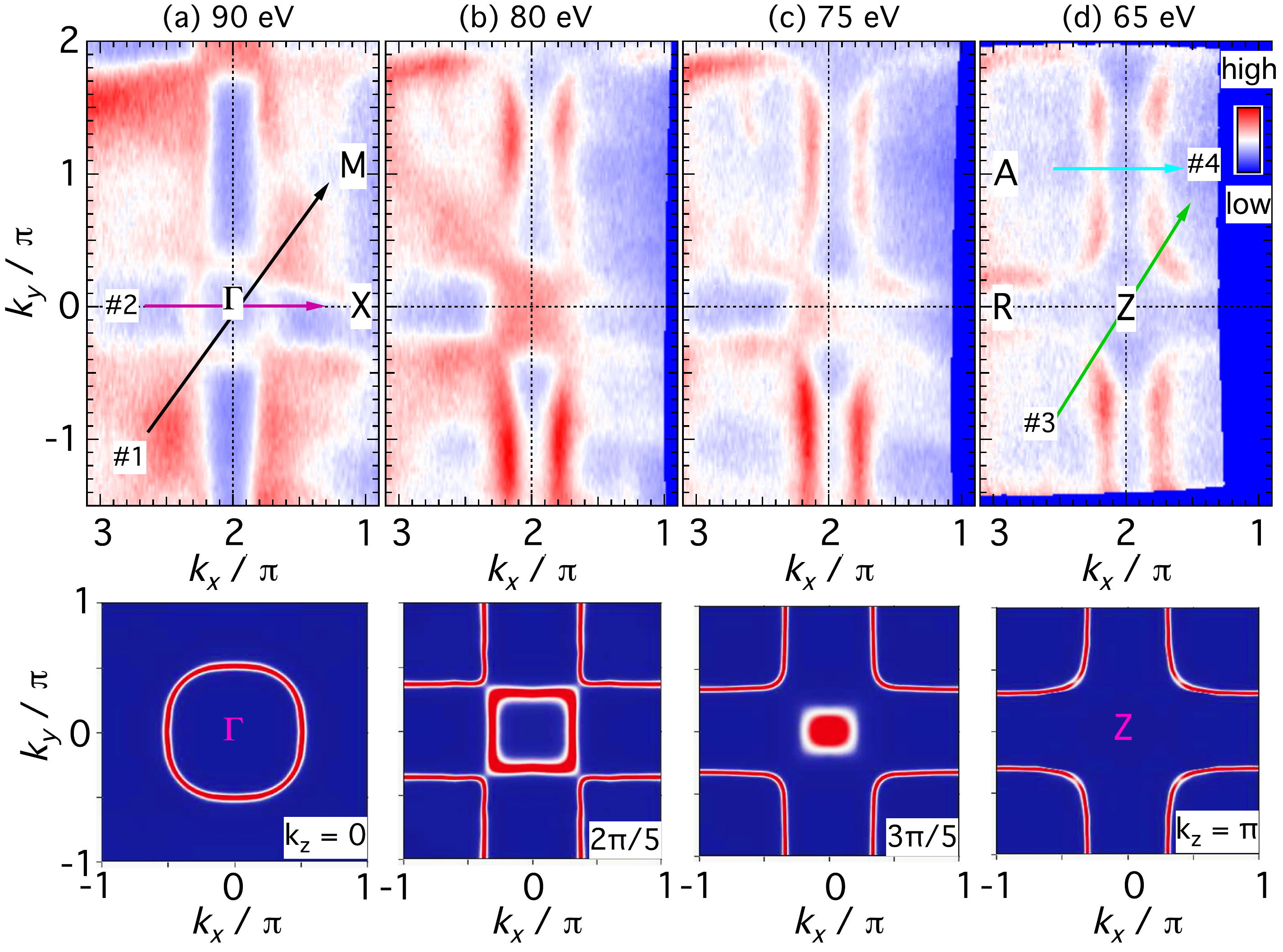}
\caption{Fermi surfaces (FSs) measured at different photon energies. (Upper panels) The FS maps of 20~uc NdNiO$_3$ thin films grown on NdGaO$_3$, measured at $T=200$~K with (a) h$\nu=90$~eV (k$_z \approx 0$), (b) 80~eV, (c) 75~eV, and (d) 65~eV (k$_z \approx \pi$). The photoemission intensity has been integrated over $\pm5$~meV. The lower panels show the calculated FSs at the roughly corresponding $k_z$ values.}
\label{fig2}
\end{figure}

\begin{SCfigure*}
\includegraphics[width=5.00in]{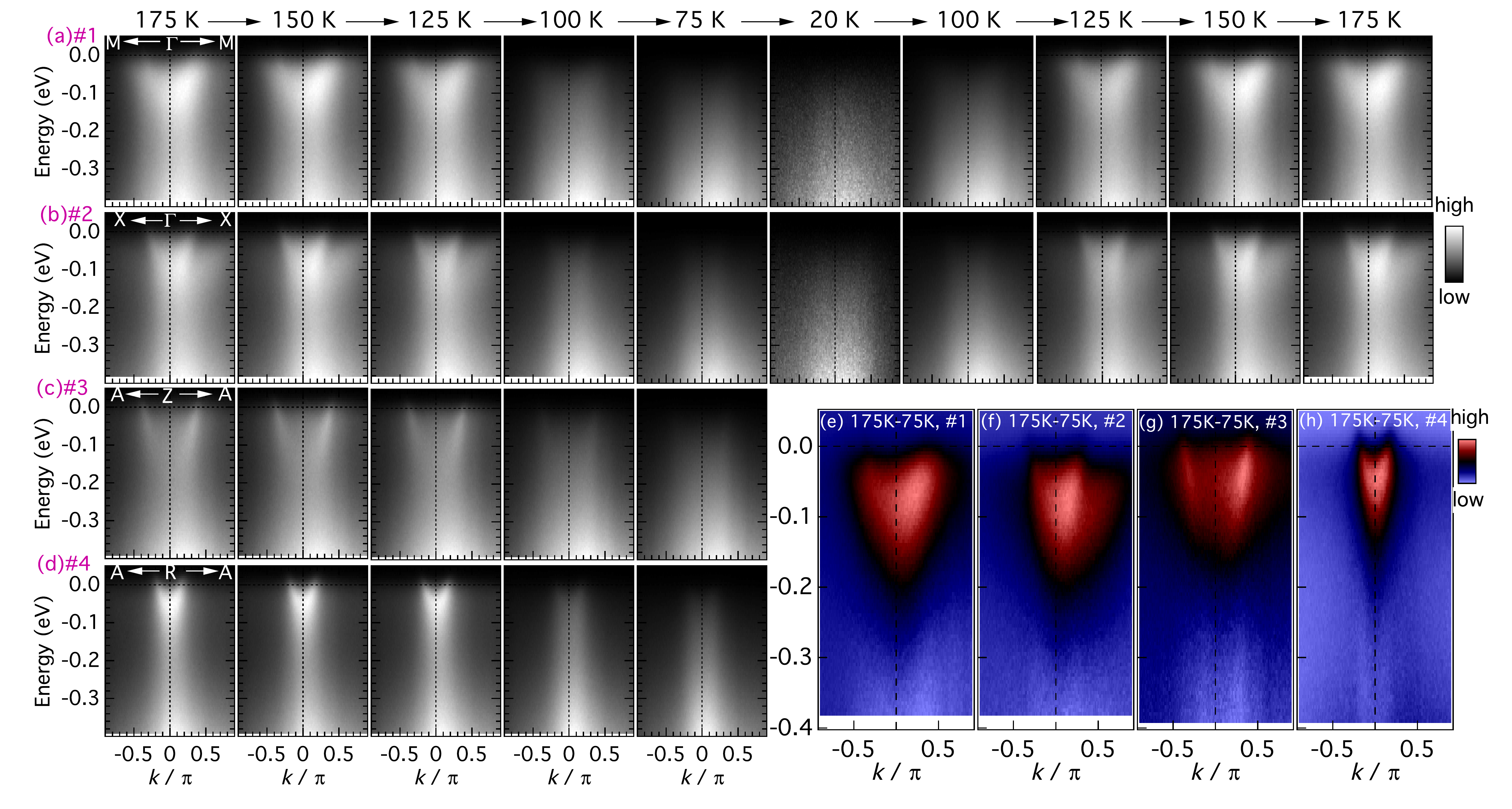}
\caption{ARPES spectra from NNO/NGO as a function of temperature. Spectra measured with h$\nu=90$~eV ($\Gamma$-point) at various sample temperatures are plotted in the upper two panels [cut\#1 ($M-\Gamma-M$) in (a) and cut\#2 ($X-\Gamma-X$) in (b)]. Corresponding spectra obtained at the $Z$-point, measured with h$\nu=65$~eV, at various sample temperatures, are plotted in the lower two panels [cut\#3 ($A-Z-A$) in (c) and cut\#4 ($A-R-A$) in (d)]. In (e--h), the difference band dispersion plots are obtained by subtracting 175~K from 75~K for all the cuts. The locations of the cuts are marked in Figs.~1(a) and 1(d).}
\label{fig3}
\end{SCfigure*}

\section{\noindent ~Results}
\subsection{\noindent I~--~NNO/NGO}

Figures~2(a)--(d) show ARPES data of NdNiO$_3$ (20~uc) grown on NdGaO$_3$(110). The maps are 2D cuts through the Fermi surface in the k$_x$-k$_y$ plane at different k$_z$ values accessed using 90, 80, 75 and 65~eV photon energies (a more complete set of maps is presented in Fig.~S4 of the Supplementary Material). The data were collected in the metallic phase at 200~K. Note that all the experimental ARPES data presented here are measured in the second Brillouin zone (BZ). At h$\nu=90$~eV [Fig.~2(a)], we observe a nearly circular FS pocket at the center of the BZ ($\Gamma$-point), which becomes smaller when viewed with photon energies of 80~eV [Fig.~2(b)] and 75~eV [Fig.~2(c)]. At the same time, clear features appear centered around the zone corners. For h$\nu=65$~eV, the photoemission intensity of the circular FS pocket at the zone center disappears completely, and a sharp band around the zone corner [$A$-point in Fig.~2(d)] forms a larger squarelike FS pocket. Therefore, our photon energy dependent measurements demonstrate the strong three-dimensional nature of these FS pockets. The measured FS topology is in good agreement with the theoretical model calculations (lower panel of Fig.~2) at the k$_z$ points corresponding with the photon energies. 

Next we show the nature of the FS pockets by studying the low energy band dispersions. From our detailed h$\nu$ dependent FS maps, we focus here on 90~eV [cut\#1 in 3(a) and cut\#2 in 3(b)] and 65~eV [cut\#3 in 3(c), cut\#4 in 3(d)] photon energies, which are close to the $\Gamma$ and $Z$-points, respectively. Figs.~3(a)--(d) show cuts along four different high-symmetry lines $M-\Gamma-M$, $X-\Gamma-X$, $A-\Gamma-A$, and $A-R-A$, respectively, which we refer to throughout the text as cuts\#1--4. In the metallic phase (175~K), the data reveal that the small pocket at the $\Gamma$-point is electron-like, while the larger pocket centred at the BZ diagonal ($A-$point) is hole-like\cite{Hansmann09,Eguchi09,Uchida11}. Theoretical calculations find that the two bands have predominantly $d_{x^2-y^2}$ and $d_{3z^2-r^2}$ characters, respectively (see Fig.~S1).

As we cool the sample, no significant change in the band dispersion and intensity is visible for temperatures down to 125~K. However, an abrupt decrease in quasiparticle spectral weight is observed when the temperature crosses below 100~K. Warming the sample reverses the behavior and restores spectral weight close to the Fermi level [Figs.~3(a,b)] following a hysteretic behavior that is also reflected in transport data [see Fig.~1(d)]. The loss of spectral weight near the Fermi level occurs for both the electron [Figs~3(a,b)] and the hole [Figs~3(c,d)] pockets. We visualize the redistribution of spectral intensity between the metallic and the insulating state by subtracting the low temperature data from the corresponding high temperature ones, as plotted in Figs.~3(e--h). The spectral weight transfer from the $E_{\rm F}$ region to higher BE is clearly visible in the entire BZ, which is found to be strongly momentum-dependent, unlike in a strongly correlated Mott scenario.

\begin{SCfigure*}
\includegraphics[width=4.90in]{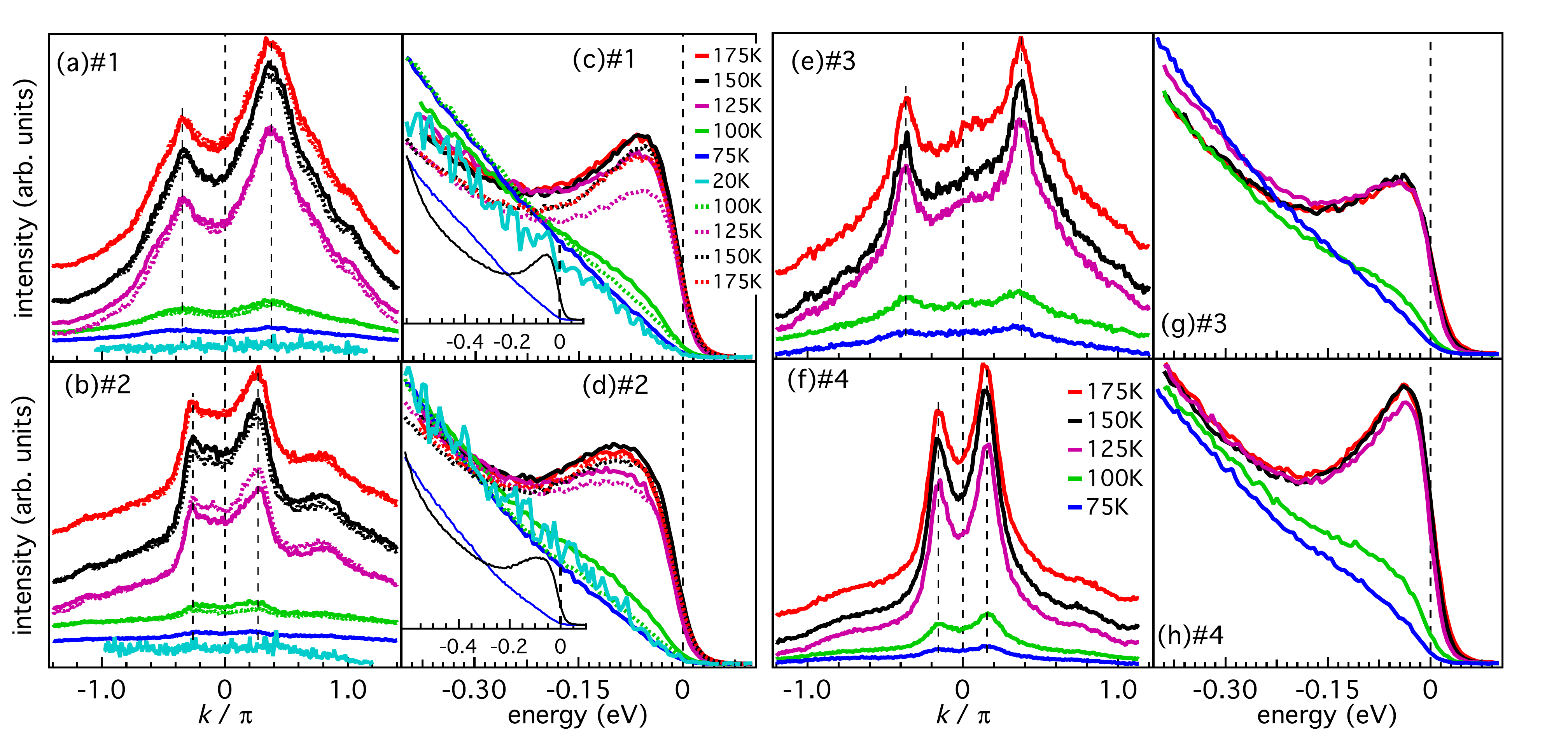}
\caption{MDCs and EDCs compared as a function of sample temperature during cooling and heating cycles between 175~K and 20~K. (a--b) MDCs at $E_{\rm F}$ and (c--d) EDCs at $k_{\rm F}$ from the spectra in Fig.~2(a,b) (h$\nu=90$~eV, $\Gamma$-point). (e--f) Analogous MDCs and (g--h) EDCs from the spectra in Fig.~2(c,d) (h$\nu=65$~eV, $Z$-point). The insets of (c) and (d) compare the EDCs over a wider energy range, revealing the spectral weight transfer from near the Fermi level to higher binding energy.}
\label{fig4}
\end{SCfigure*}

In Fig.~4 we provide more comprehensive analysis of the spectral weight transfer in the momentum distribution curves (MDCs) and energy distribution curves (EDCs) at different temperatures. In Figs.~4(a,b) and 4(e,f), the MDCs of electron and hole bands extracted from the panels (a,b) and (c,d) of Fig.~3 are presented. The corresponding EDCs at the Fermi momentum (k$_F$) are shown in Figs.~4(c,d) and 4(g,h). Note that the data points shown earlier in Fig.~1(d) were obtained by integrating the EDCs in Fig.~4(c) from -0.2 to 0.1 eV. These show a step-like, hysteretic change in spectral intensity centred near the transport-measured transition temperature of about 140~K. All MDCs at 175~K exhibit two well-defined peaks with substantial spectral weight at the Fermi level. There is virtually no change in the MDC peak positions as the temperature is reduced; however, the peak intensity is significantly suppressed below 125~K relative to higher temperatures. In all the cuts, the EDCs at 175~K show a high intensity quasiparticle peak extending from the Fermi level up to $E_{\rm F}-0.2$~eV. There is no drastic change upon lowering the temperature up to 125~K; however, a significant decrease in intensity is observed at around 100~K. The quasiparticle spectral weight continues to decrease as the temperature is lowered to about 75~K after which point the intensity loss appears to saturate (i. e., the EDCs for 75~K and 20~K are almost overlapping). Most intriguingly, the intensity below 0.2~eV increases with decreasing temperature. Earlier angle-integrated photoemission studies of Nd$_{1-x}$Sm$_x$NiO$_3$ show similar behavior of the spectral weight transfer from near $E_{\rm F}$ region to higher binding energy region \cite{Okazaki03}. Our observation is also consistent with an optical study, where the MIT is argued to be linked to a spectral weight transfer from the Drude region (coherent state) to the deeper energy\cite{StewartPRL11,Stewart12}.

\subsection{\noindent II~--~NNO/LAO}

In order to gain insight into the origins of the MIT in NNO and the accompanying spectroscopic changes seen by ARPES, we have leveraged the natural advantages of thin films as tunable systems, exploiting, in particular, different substrates to change the epitaxial strain and thereby modify the electronic properties of NNO. Earlier studies have shown that substrate-induced strain alters the MIT, but the specific mechanism by which it influences the MIT has remained unknown. The temperature dependence of the resistivity of NNO/LAO(001), where the strain is compressive ($\approx-0.3\%$), is notably different from that of the tensile-strained ($\approx+1.4\%$) NNO/NGO(110) films studied above. Namely, our 20-uc thick NNO/LAO films show metallic behavior down to 60~K with broad hysteresis in the MIT that is not saturated down to the lowest measurement temperature of 2~K, whereas the NNO/NGO samples have an MIT temperature of about 140~K and a narrower hysteresis of about 30~K [see Fig.~1(d)]. While the strain-dependent behavior of the MIT in NNO films varies somewhat among recent studies, the reduced transition temperature we observe in NNO/LAO is in qualitative agreement with recent transport results \cite{Scherwitzl}. Another study found that the temperature-driven MIT can even be completely suppressed under compressive strain, which appears to coincide with the disappearance of antiferromagnetic ordering that otherwise is found in tensile-strained NNO \cite{Liu13}. These observations suggest a unique role of the compressive strain in suppressing the insulating phase, as well as magnetic ordering at low temperature.

Reflecting their different electronic behavior, ARPES reveals marked differences between the electronic structures of tensile-strained NNO/NGO and compressively-strained NNO/LAO. Figs~5(a--d) show FS maps measured with 90~eV ($\Gamma$-point) and 65~eV ($Z$-point) photon energies, which show the persistence of the metallic FS down to 20~K. The overall shape of the FS remains similar at high and low temperatures, and the quasiparticles--which are still evident at 20~K---sharpen as the temperature is decreased. Resistance measurements of our NNO/LAO also show the absence of a complete insulating phase down to 2~K [Fig.~1(d)]. At 200~K, the FS of NNO/LAO contains an electron pocket centered at $\Gamma$ and cuboidal hole pockets centered at the BZ corners. These resemble the Fermi surface components of NNO/NGO, except that the hole pockets in NNO/LAO are noticeably larger and the electron pocket is slightly smaller. Strikingly, however, in NNO/LAO, extra pockets appear at the corners of the BZ [see around $A$-point at k$_x$, k$_y$ $\approx$ 3, -1 and 3, 1 in Figs.~5(c,d)]. Such bands are not visible in NNO/NGO within the binding energy range extending $\sim$300~meV from $E_F$ that we have studied. This observation suggests that the strain not only causes band renormalization  as one would anticipate in general, but it also changes the band ordering. With theoretical modeling we find that strain also changes the crystal field splitting, pushing the $t_{2g}$ orbitals upward, while pulling the $e_g$ orbitals further down. This makes the electron-pocket at the $\Gamma$ pocket shrink, while the hole-pocket at the Z point grows [as shown in Fig.~5(g,h) by extracting the peak positions of MDCs at each energy and in Figs.~5(i,j) by the theoretical calculations along cut\#4 ($A-R-A$)], and additionally a tiny hole pocket also forms at $(\pi,\pi,\pi)$ point as the $d_{xy}$ orbitals now cross the Fermi level in NNO/LAO system.

\begin{figure}
\includegraphics[width=3.2in]{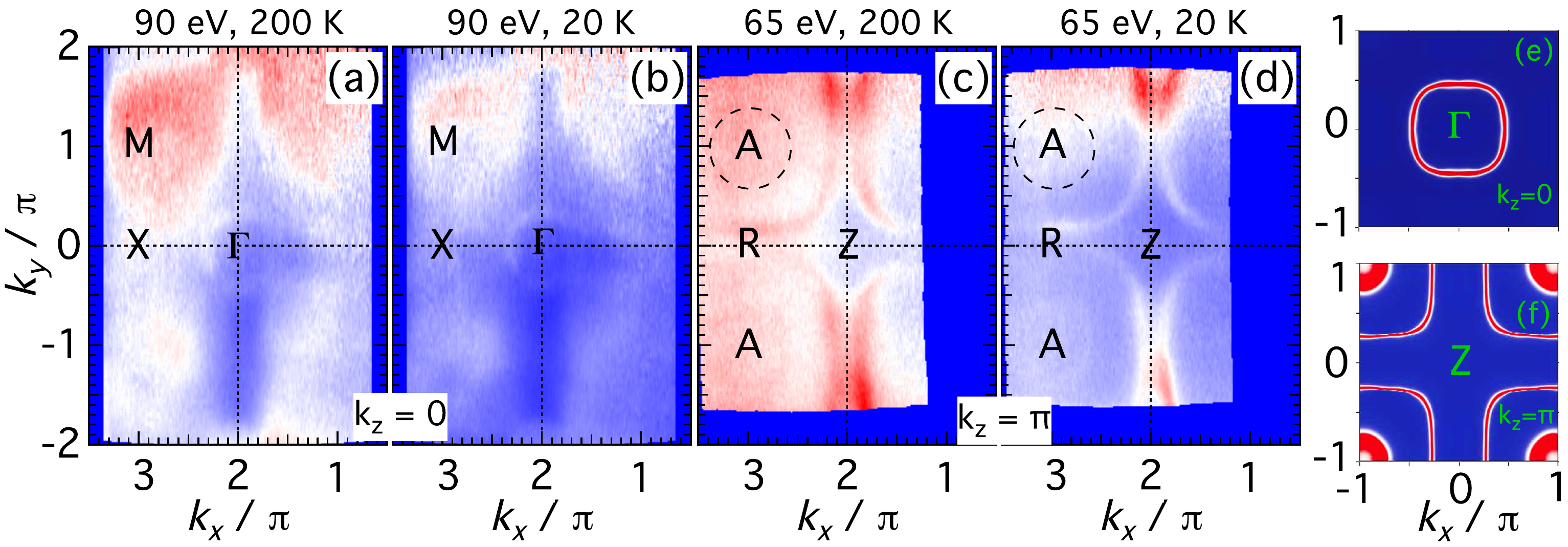}
\includegraphics[width=3.2in]{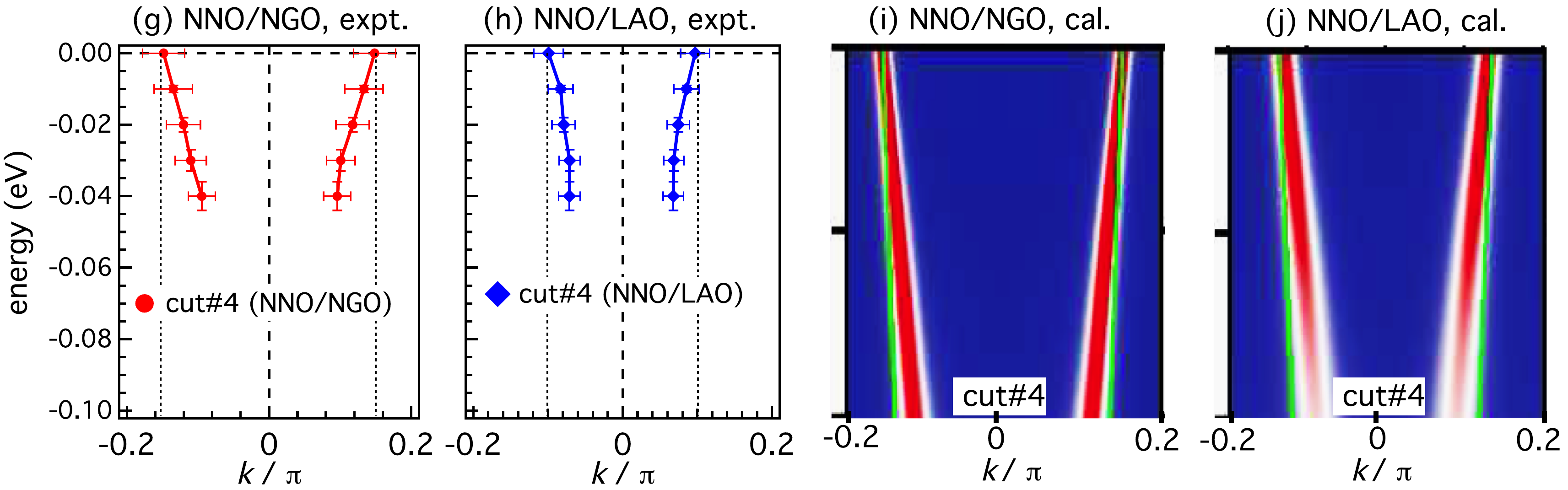}
\caption{(a-d) Fermi surface maps of 20~uc NdNiO$_3$ thin films grown on LaAlO$_3$, measured at 200~K and 20~K using h$\nu=90$~eV, and 65~eV. In order to make the extra holelike pocket (around $A$-point) clearly visible, we have marked the position by dotted circle around $A$-point in the upper corner of (c,d). Theoretically calculated FSs are plotted in (e,f). (g,h) Band dispersions of NNO/NGO and NNO/LAO and (i, j) plots of the theoretically calculated ARPES spectral function (including self-energy) for the same cut\#4. The corresponding non-interacting bands are overlaid in (i,j).}
\label{fig4}
\end{figure}

\begin{figure*}
\includegraphics[width=5.95in]{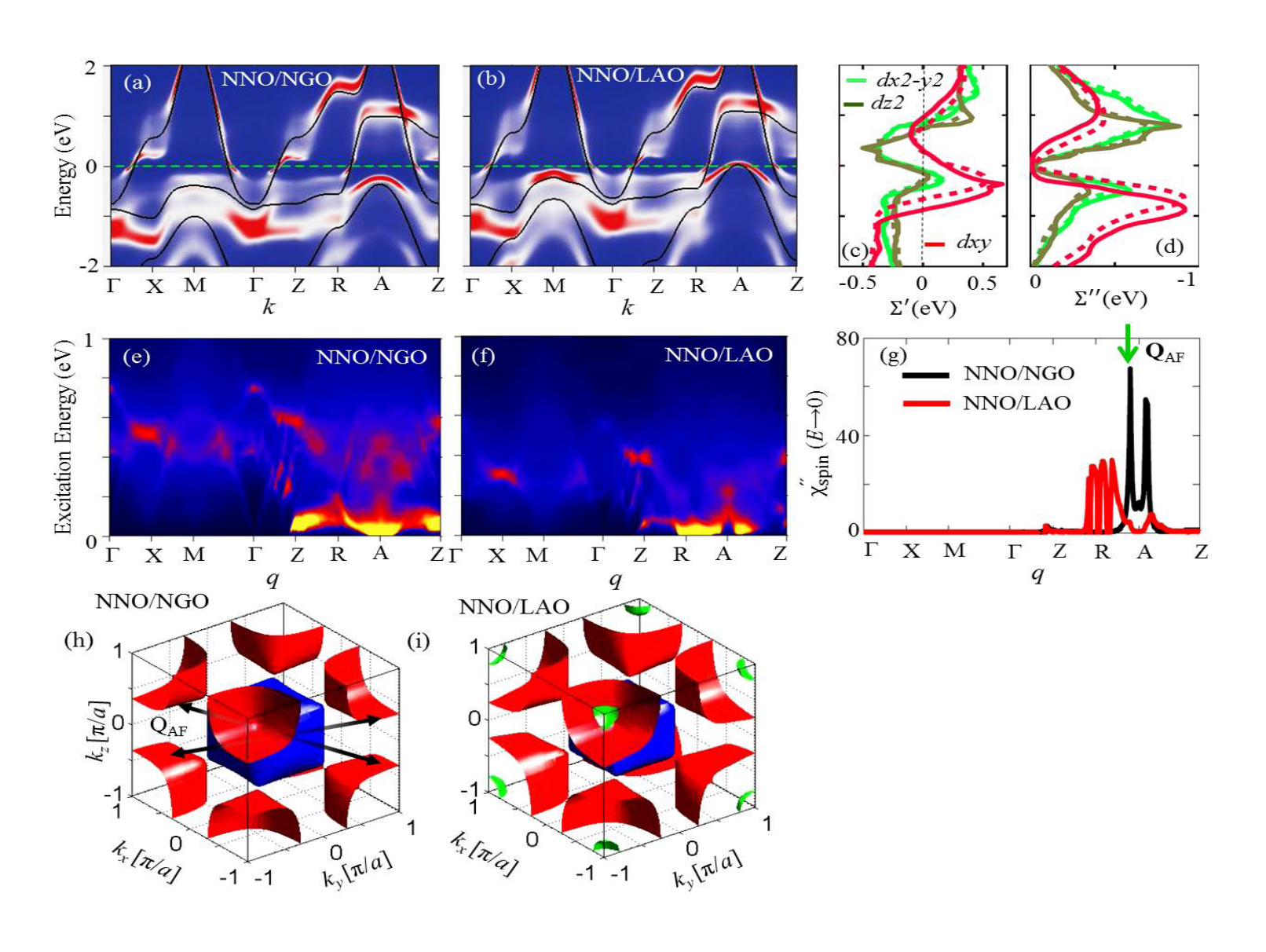}
\caption{Theoretical results of the spectral weight renormalization and self-energy of NNO/NGO and NNO/LAO. (a,b) Spectral weight maps along high-symmetry momentum cuts, computed with the inclusion of a many-body $\Sigma({\bf k},E)$ due to dynamical electron-electron interactions, and compared with the corresponding non-interacting band structure, plotted in black solid line. (c, d) Real and imaginary parts of self-energy at ${\bf k}=0,0,0$ for both NNO/NGO (solid lines) and NNO/LAO (dashed lines) systems. (e, f) Full spectrum of many-body spin-correlation function. (g) susceptibility compared in the low-energy region. (h,i) calculated Fermi surfaces for both the systems.}
\label{fig6}
\end{figure*}

\section{\noindent ~Discussion}
\vskip -0.4cm
Our ARPES experiments reveal two important observations related to the MIT. Firstly, the spectral weight transfer across the MIT from the quasiparticle states near Fermi level to the high-energy incoherent states is strongly anisotropic, indicating that the correlation effect is momentum dependent. Secondly, the substrate driven strain changes the FS areas, and thereby controls the FS nesting driven insulating gap. In order to understand the mechanism of MIT, we have combined our results with the theory.  
The momentum-dependent band renormalization and spectral weight transfer is well captured within the MRDF theory, as demonstrated in Fig.~6 (for more detail, see the supplementary information). The model captures the intermediate coupling regime of the interaction, in which the effective interaction is of the order of the bandwidth \cite{AIPDas}. In this limit, the electronic spectrum is often split into low-energy itinerant and high-energy localized states by coupling to spin, charge and other density fluctuations. The added benefit of the MRDF model is that it captures the full momentum and energy dependence of the self-energy, which smoothly interpolate the band and spectral weight between the itinerant and localized states. 

The calculated self-energy-dressed spectral weight maps are presented in Figs.~6(a) and 6(b), and compared with the corresponding tight-binding band structure (solid lines) for NNO/NGO and NNO/NGO, respectively. Considerable deviation of the calculated spectral function with respect to its non-interacting band is visible here. Both real ($\Sigma^{\prime}$) and imaginary ($\Sigma^{\prime\prime}$) parts of the spectral function (which are related to the band renormalization and lifetime, respectively) are plotted in Fig.~6(c,d) at $k=0,0,0$ for both NNO/NGO (solid lines) and NNO/LAO (dashed lines), while its momentum dependence is presented in the supplementary information. For the NNO/NGO, we notice that the self-energies have higher intensity, and the corresponding slope in $\Sigma^{\prime}$ (determining the strength of mass enhancement at the Fermi level) is also larger than the case of NNO/LAO. $\Sigma^{\prime}(E<0)$ shows a peak around 0.2-0.3~eV, while $\Sigma^{\prime\prime}$ shows a peak where $\Sigma^{\prime}$ changes sign. These calculated results \cite{AIPDas,PuDas,MoS2Das} are in good agreement with our ARPES experimental observation of the crossover that occurs at $E_{\rm F}-0.2$~eV between the strongly renormalized band near $E_{\rm F}$ and the localized like states at higher energy. The computed band renormalization factor $Z_{\bf k}=(1-\partial \Sigma^{\prime}({\bf k},\omega)/\partial \omega)_{\omega=0}^{-1}$ changes characteristically throughout the BZ, and is found to be stronger in NNO/NGO than NNO/LAO, as discussed in the supplementary information. 
 
Secondly, our ARPES data demonstrated that the FS areas changes between NNO/NGO and NNO/LAO. In addition, for the NNO/LAO system, a band crosses $E_{\rm F}$ at the $A$-point, forming the extra hole like pockets [also can be seen in the calculated FSs as shown in Fig.6(i)]. In order to explain this finding, we model the crystal field splitting by parameterizing the relative onsite potential between the low-lying $e_{g}$ and $t_{2g}$ orbitals, as well as the hybridization between them. These two parameters lower the chemical potential for the $e_{g}$ orbitals, while lifting the $t_{2g}$ orbital closer to the Fermi level in NNO/LAO, as seen experimentally. Recently published x-ray absorption spectra also show a remarkably large and asymmetric energy shift in the O K-edge as a function of strain \cite{Liu13}, suggesting the same conclusion. H. K. Yoo, {\it et al.} observed the strain dependent FS superstructures in LaNiO$_3$ films \cite{Yoo14}, which may be due to the modification in the low-lying orbitals and crystal field splitting. It has been suggested that the existence of charge disproportionation in RNiO$_3$ is strongly related to the metal-insulator phase transition \cite{StaubPRL02,Scagnoli05,MedardePRB09,Bilewska10}. Our ARPES results on two differently strained films suggest the existence of a tunable electronic localization. If so, then the number of itinerant charges, above the transition in the metallic state can be different for NNO films grown on NGO and LAO. Indeed, we find that strain-tuned crystal field splitting alters the FS nesting conditions, in turn influencing the spin fluctuations that play a dominant role in the MIT.

We further performed the calculations in order to capture the full spectrum of the spin susceptibility for the strained NNO films, see Figs.~6(e,f). Notably, we find that the spin susceptibility dominates over the charge susceptibility in both systems. The RPA susceptibilites exhibit two dominant intensities at different energy and momentum scales. The low-energy collective modes, such as magnon with preferred wavevector indicate the presence of electronic or magnetic order, while the high-energy dispersive modes, like paramagnons, gives the marginal Fermi-liquid-like state. The spin susceptibility, in the low-energy region, is stronger in NNO/NGO and concentrated around the wavevector Q$_{AF}$(1/4, 1/4, 1/4$\pm$$\delta$)$\pi/a$, while for NNO/LAO, it is spread out over a large momentum region, as shown in Fig.~6(e--g). The origin of the stronger spin-fluctuation spectral weight at one dominant Q$_{AF}$ value in NNO/NGO [Fig.~6(g)] is related to the underlying band structure and nesting-driven FS instability, which is found to be weaker in NNO/LAO. This indicates that NNO/NGO is prone to a magnetic instability through dynamical fluctuations resulting in its stronger band renormalization found in the metallic state, which ultimately may account for its higher MIT.

\section{\noindent ~Conclusion}

In summary, we have presented the first momentum-resolved study of the electronic structure of NNO thin films across the MIT as a function of epitaxial strain (using two different substrates). Both compressive and tensile strained films in the metallic phase showed 3D FS and share common elements: spheroidal electron like pockets centered at $\Gamma$ and large cuboidal hole like pockets at the 3D BZ corners. The corresponding bands are strongly renormalized in the metallic phase, in agreement with theoretical calculations. A transfer of the spectral weight from near $E_{\rm F}$ region to higher binding energy occurs in the entire BZ across the MIT. On the basis of these observations, we have performed a self-energy calculation due to spin and charge fluctuation in the dynamical correlation limit, which shows that the intertwined FS nesting and the dynamical correlation effects are stronger in NNO/NGO than NNO/LAO. We find that the strain reduces crystal field splitting in NNO/LAO, lowering the $e_{g}$ states and lifting those of the $t_{2g}$ orbitals. This  leads to the appearance of a new holelike  FS at the $A-$point, which plays a crucial role in lowering the MIT. Our study provides insights into the physical mechanisms responsible of driving the MIT in complex oxides, implying how strain, nesting, correlations, and magnetic fluctuations can be tuned, which may help the future development of rare earth nickelates in applications \cite{Mannhart,Yang11,Scherwitzl,Shi13}.

\section {Acknowledgment}
The authors thank technical staff at the SIS beam line for their cooperation and help during the experiments.


\begin{thebibliography}{99}

\bibitem{Imada98} M. Imada, A. Fujimori, and Y. Tokura, Metal-insulator transitions, Rev. Mod. Phys. {\bf 70}, 1039 (1998).

\bibitem{HwangNM12} H. Y. Hwang, Y. Iwasa, M. Kawasaki, B. Keimer, N. Nagaosa, and Y. Tokura, Emergent phenomena at oxide interfaces, Nature Materials {\bf 11}, 103 (2012). 

\bibitem{Zubko} P. Zubko, S. Gariglio, M. Gabay, P. Ghosez, J. M. Triscone, Interface physics in complex oxide heterostructures, Annu. Rev. Condens. Matter Phys. {\bf 2}, 141 (2011). 

\bibitem{Medarde97} M. Medarde, Structural, magnetic and electronic properties of RNiO$_3$ perovskites (R = rare earth), J. Phys.: Condens. Matter {\bf 9}, 1679 (1997).

\bibitem{Catalan} G. Catalan, Progress in perovskite nickelate research, Phase Transitions {\bf 81}, 729 (2008).

\bibitem{Liu13} J. Liu, M. Kargarian, M. Kareev, B. Gray, P. J. Ryan, A. Cruz, N. Tahir, Yi-De Chuang, J. Guo, J. M. Rondinelli, J. W. Freeland, G. A. Fiete, and J. Chakhalian, Heterointerface engineered electronic and magnetic phases of NdNiO$_3$ thin films, Nature Communications {\bf 4}, 2714 (2013). 

\bibitem{LiuPRL12} J. Liu, M. Kareev, D. Meyers, B. Gray, P. Ryan, J. W. Freeland, and J. Chakhalian, Metal-insulator transition and orbital reconstruction in Mott-type quantum wells made of NdNiO$_3$, Phys. Rev. Lett. {\bf 109}, 107402 (2012).

\bibitem{ScherwitzlPRL} R. Scherwitzl, S. Gariglio, M. Gabay, P. Zubko, M. Gibert, and J.-M. Triscone,
Metal-insulator transition in ultrathin LaNiO$_3$ films, Phys. Rev. Lett. {\bf 106}, 246403 (2011).

\bibitem{Scherwitzl} R. Scherwitzl, P. Zubko, I. G. Lezama, S. Ono, A. F. Morpurgo, G. Catalan, and J. M. Triscone, Electric-field control of the metal-insulator transition in ultrathin NdNiO$_3$ films, Advanced Materials {\bf 22}, 5517 (2010). 

\bibitem{CatalanPRB00} G. Catalan, R. M. Bowman, and J. M. Gregg, Metal-insulator transitions in NdNiO$_3$ thin films, Phys. Rev. B {\bf 62}, 7892  (2000).

\bibitem{Stewart12} M. K. Stewart, D. Brownstead, J. Liu, M. Kareev, J. Chakhalian, and D. N. Basov, Heterostructuring and strain effects on the infrared optical properties of nickelates, Phys. Rev. B {\bf 86}, 205102 (2012).

\bibitem{Benckiser11} E. Benckiser {\it et al.}, Orbital reflectometry of oxide heterostructures. Nat. Materials {\bf 10}, 189 (2011).

\bibitem{Wu13} M. Wu {\it et al.}, Strain and composition dependence of orbital polarization in nickel oxide superlattices. Phys. Rev. B {\bf 88}, 125124 (2013). 

\bibitem{Canfiled} P. C. Canfield, J. D. Thompson, S. W. Cheong, and L. W. Rupp, Extraordinary pressure dependence of the metal-to-insulator transition in the charge-transfer compounds NdNiO$_3$ and PrNiO$_3$, Phys. Rev. B {\bf 47}, 12357 (1993).

\bibitem{Obradors} X. Obradors, L. M. Paulius, M. B. Maple, J. B. Torrance, A. I. Nazzal, J. Fontcuberta, and X. Granados, Pressure-dependence of the metal-insulator-transition in the charge-transfer oxides RNiO$_3$ (R = Pr, Nd, Nd$_{0.7}$La$_{0.3}$), Phys. Rev. B {\bf 47}, 12353 (1993).


\bibitem{Kaiser11} A. M. Kaiser, A. X. Gray, G. Conti, J. Son, A. Greer, A. Perona, A. Rattanachata, A.Y. Saw, A. Bostwick, S. Yang, S.-H. Yang, E. M. Gullikson, J. B. Kortright, S. Stemmer, and C. S. Fadley, Suppression of near-Fermi level electronic states at the interface in a LaNiO$_3$/SrTiO$_3$ superlattice, Phys. Rev. Lett. {\bf 107}, 116402 (2011).

\bibitem{Munoz09} J. L. Garc\'{i}a-Mu\~{n}oz, M. A. G. Aranda, J. A. Alonso, and M. J. Mart\'{i}nez-Lope, Structure and charge order in the antiferromagnetic band-insulating phase of NdNiO$_3$, Phys. Rev. B {\bf 79}, 134432 (2009).

\bibitem{Scagnoli08} V. Scagnoli, U. Staub, Y. Bodenthin, M. Garc'{i}a-Fern'{a}ndez, A. M. Mulders, G. I. Meijer, and G. Hammerl, Induced noncollinear magnetic order of Nd$^{3+}$ in NdNiO$_3$ observed by resonant soft x-ray diffraction, Phys. Rev. B {\bf 77}, 115138 (2008).

\bibitem{MedardePRB92} M. Medarde, A. Fontaine, J. L. Garc\'{i}a-Mu\~{n}oz, J. Rodr�guez-Carvajal, M. de Santis, M. Sacchi, G. Rossi, and P. Lacorre, RNiO$_3$ perovskites (R = Pr, Nd): Nickel valence and the metal-insulator transition investigated by x-ray absorption spectroscopy, Phys. Rev. B {\bf 46}, 14975 (1992).

\bibitem{Kumar13} D. Kumar, K. P. Rajeev, J. A. Alonso, and M. J. Mart\'{i}nez-Lope, Spin-canted magnetism and decoupling of charge and spin ordering in NdNiO$_3$, Phys. Rev. B {\bf 88}, 014410 (2013).

\bibitem{Scagnoli06} V. Scagnoli, U. Staub, A. M. Mulders, M. Janousch, G. I. Meijer, G. Hammerl, J. M. Tonnerre, and N. Stojic, Role of magnetic and orbital ordering at the metal-insulator transition in NdNiO$_3$, Phys. Rev. B {\bf 73}, 100409(R) (2006).

\bibitem{HanPRL11} M. J. Han, X. Wang, C. A. Marianetti, and A. J. Millis, Dynamical mean-field theory of nickelate superlattices, Phys. Rev. Lett. {\bf 107}, 206804 (2011).

\bibitem{ParkPRL12} H. Park, A. J. Millis, and C. A. Marianetti, Site-selective Mott transition in rare-earth-element nickelates, Phys. Rev. Lett. {\bf 109}, 156402 (2012).

\bibitem{LauPRL13} B. Lau,  and A. J. Millis, Theory of the magnetic and metal-insulator transitions in RNiO$_3$ bulk and layered structures, Phys. Rev. Lett. {\bf 110}, 126404 (2013).

\bibitem{Johnston} S. Johnston, A. Mukherjee, I. Elfimov, M. Berciu, and G. A. Sawatzky, Charge disproportionation without charge transfer in the rare-earth-element nickelates as a possible mechanism for the metal-insulator transition, Phys. Rev. Lett. {\bf 112}, 106404 (2014).

\bibitem{StaubPRL02} U. Staub, G. I. Meijer, F. Fauth, R. Allenspach, J. G. Bednorz, J. Karpinski, S. M.  Kazakov, L. Paolasini, and F. d'Acapito, Direct observation of charge order in an epitaxial NdNiO$_3$ film, Phys. Rev. Lett. {\bf 88}, 126402 (2002).

\bibitem{Scagnoli05} V. Scagnoli, U. Staub, M. Janousch, A. M. Mulders, M. Shi, G. I. Meijer, S. Rosenkranz, S. B. Wilkins, L. Paolasini, J. Karpinski, S. M. Kazakov, and S. W. Lovesey, Charge disproportionation and search for orbital ordering in NdNiO$_3$ by use of resonant x-ray diffraction, Phys. Rev. B {\bf 72}, 155111 (2005).

\bibitem{MedardePRB09} M. Medarde, C. Dallera, M. Grioni, B. Delley, F. Vernay, J. Mesot, M. Sikora, J. A.  Alonso, and M. J. Martinez-Lope, Charge disproportionation in RNiO$_3$ perovskites (R = rare earth) from high-resolution x-ray absorption spectroscopy, Phys. Rev. B {\bf 80}, 245105 (2009).

\bibitem{Bilewska10} K. Bilewska, E. Wolna, M. Edely, P. Ruello, and J. Szade, Evidence of charge disproportionation on the nickel sublattice in EuNiO$_3$ thin films: X-ray photoemission studies, Phys. Rev. B {\bf 82}, 165105 (2010).

\bibitem{MedardePRL98} M. Medarde, P. Lacorre, K. Conder, F. Fauth, and A. Furrer, Giant $^{16}$O--$^{18}$O isotope effect on the metal-insulator transition of RNiO$_3$ perovskites ( R = Rare Earth), Phys. Rev. Lett. {\bf 80}, 2397 (1998).

\bibitem{StewartPRL11} M. K. Stewart, J. Liu, M. Kareev, J. Chakhalian, and D. N. Basov, Mott physics near the insulator-to-metal transition in NdNiO$_3$, Phys. Rev. Lett. {\bf 107}, 176401 (2011).

\bibitem{Yoo14} H. K. Yoo, {\it et al.}, 
Dimensional crossover of the electronic structure in LaNiO$_3$ ultrathin films: Orbital reconstruction, Fermi surface nesting, and the origin of the metal-insulator transition, arXiv:1309.0710; Latent instabilities in metallic LaNiO$_3$ films by strain control of Fermi-surface topology, Scientific Reports {\bf 5}, 8746 (2015).

\bibitem{King14} P. D. C. King, H. I. Wei, Y. F. Nie, M. Uchida, C. Adamo, S. Zhu, X. He, I. Bo\v{z}ovi\'{c}, D. G. Schlom and K. M. Shen, Atomic-scale control of competing electronic phases in ultrathin LaNiO$_3$, Nature Nanotechnology {\bf 9}, 443 (2014).

\bibitem{Hepting} M. Hepting, M. Minola, A. Frano, G. Cristiani, G. Logvenov, E. Schierle, M. Wu, M. Bluschke, E. Weschke, H.-U. Habermeier, E. Benckiser, M. Le Tacon, and B. Keimer, Tunable charge and spin order in PrNiO$_3$ thin films and superlattices, Phys. Rev. Lett. {\bf 113}, 227206 (2014).

\bibitem{Boris} A. V. Boris {\it et al.}, Dimensionality Control of Electronic Phase Transitions in Nickel-Oxide Superlattices. Science {\bf 332}, 937 (2011). 

\bibitem{Okazaki03} K. Okazaki, T. Mizokawa, A. Fujimori, E. V. Sampathkumaran, M. J. Martinez-Lope, and J. A. Alonso, Crossover in the nature of the metallic phases in the perovskite-type RNiO$_3$, Phys. Rev. B {\bf 67}, 073101 (2003).

\bibitem{Schwier12} E. F. Schwier, R. Scherwitzl, Z. Vydrova, M. Garcia-Fernandez, M. Gibert, P. Zubko, M. G. Garnier, J.-M. Triscone, and P. Aebi, Unusual temperature dependence of the spectral weight near the Fermi level of NdNiO$_3$ thin films, Phys. Rev. B {\bf 86}, 195147 (2012).

\bibitem{Vobornik99} I. Vobornik, L. Perfetti, M. Zacchigna, M. Grioni, G. Margaritondo, J. Mesot, M. Medarde, and P. Lacorre, Electronic-structure evolution through the metal-insulator transition in RNiO$_3$, Phys. Rev. B {\bf 60}, 8426(R) (1999).

\bibitem{Barman} S. R. Barman, A. Chainani, and D. D. Sarma, Covalency-driven unusual metal-insulator transition in nickelates, Phys. Rev. B {\bf 49}, 8475 (1994).

\bibitem{MedardeEPL97} M. Medarde, D. Purdie, M. Grioni, M. Hengsberger, Y. Baer, and P. Lacorre, A photoemission spectroscopy study of PrNiO$_3$ through the metal-insulator transition, Europhy. Lett., {\bf 37}, 483 (1997).

\bibitem{Sassa} Y. Sassa, M. Radovi{\'c}, M. Mansson, E. Razzoli, X. Y. Cui, S. Pailh{\`e}s, S. Guerrero, M. Shi, P. R. Willmott, F. Miletto Granozio, J. Mesot, M. R. Norman, and L. Patthey, Ortho-II band folding in YBa$_2$Cu$_3$O$_{7-\delta}$ films revealed by angle-resolved photoemission, Phys. Rev. B {\bf 83}, 140511(R) (2011). 

\bibitem{AIPDas} T. Das, R. S. Markiewicz, and A. Bansil, Intermediate coupling model of the cuprates, Adv. Phys. {\bf 63}, 151-266 (2014). 

\bibitem{PuDas} T. Das, J. -X. Zhu, and M. J. Graf, Spin-fluctuations and the peak-dip-hump feature in the photoemission spectrum of actinides, Phys. Rev. Lett. {\bf 108}, 017001 (2012). 

\bibitem{MoS2Das} T. Das, and K. Dolui, Superconducting dome in MoS2 and TiSe2 generated by quasiparticle-phonon coupling,  Phys. Rev. B {\bf 91}, 094510 (2015). 

\bibitem{Hansmann09} P. Hansmann, X. Yang, A. Toschi, G. Khaliullin, O. K. Andersen, and K. Held, Turning a nickelate Fermi surface into a cupratelike one through heterostructuring, Phys. Rev. Lett. {\bf 103}, 016401 (2009).

\bibitem{Eguchi09} R. Eguchi, A. Chainani, M. Taguchi, M. Matsunami, Y. Ishida, K. Horiba, Y. Senba, H. Ohashi, and S. Shin, Fermi surfaces, electron-hole asymmetry, and correlation kink in a three-dimensional Fermi liquid LaNiO$_3$, Phys. Rev. B {\bf 79}, 115122 (2009).

\bibitem{Uchida11} M. Uchida, K. Ishizaka, P. Hansmann, Y. Kaneko, Y. Ishida, X. Yang, R. Kumai, A. Toschi, Y. Onose, R. Arita, K. Held, O. K. Andersen, S. Shin, and Y. Tokura, Pseudogap of metallic layered nickelate R$_{2-x}$Sr$_x$NiO$_4$ (R = Nd, Eu) crystals measured using angle-resolved photoemission spectroscopy, Phys. Rev. Lett. {\bf 106}, 027001 (2011).

\bibitem{Mannhart} J. Mannhart, and D. G. Schlom, Oxide interfaces--An opportunity for electronics, Science {\bf 327}, 1607 (2010). 

\bibitem{Yang11} Z. Yang, C. Ko, and S. Ramanathan, Oxide electronics utilizing ultrafast metal-insulator transitions, Annu. Rev. Condens. Matter Phys. {\bf 41}, 337 (2011). 

\bibitem{Shi13} J. Shi, S. D. Ha, Y. Zhou, F. Schoofs, and S. Ramanathan, A correlated nickelate synaptic transistor, Nature Communications {\bf 4}, 2676 (2013). 

\end{thebibliography}
\end{document}